\documentclass[11pt]{article}

\usepackage{amsmath}
\usepackage{graphicx}
\usepackage{amsfonts}
\usepackage{amssymb}
\usepackage{epsfig}
\usepackage{color}
\usepackage{psfrag}
\usepackage{epstopdf}
\usepackage{caption}
\usepackage{subcaption}

\usepackage{adjustbox}

\newcommand{\EQ}{\begin{equation}}
\newcommand{\EN}{\end{equation}}
\newcommand{\be}{\begin{equation}}
\newcommand{\ee}{\end{equation}}
\newcommand{\bea}{\begin{eqnarray}}
\newcommand{\eea}{\end{eqnarray}}

\begin{document} \setcounter{page}{0}
\newpage
\setcounter{page}{0}
\renewcommand{\thefootnote}{\arabic{footnote}}
\newpage
\begin{titlepage}
\begin{flushright}
\end{flushright}
\vspace{0.5cm}
\begin{center}
{\large {\bf Criticality in the disordered $N$-color Ashkin-Teller model}}\\
\vspace{1.8cm}
{\large Youssef Makoudi$^{1,2}$ and Gesualdo Delfino$^{1,2}$}\\
\vspace{0.5cm}
{\em $^1$SISSA -- Via Bonomea 265, 34136 Trieste, Italy}\\
{\em $^2$INFN sezione di Trieste, 34100 Trieste, Italy}\\

\end{center}
\vspace{1.2cm}

\renewcommand{\thefootnote}{\arabic{footnote}}
\setcounter{footnote}{0}

\begin{abstract}
\noindent
The $N$-color Ashkin-Teller model corresponds to $N$ Ising models coupled by four-spin interactions. We consider the two-dimensional case in presence of quenched disorder and use scale invariant scattering theory to determine all the solutions of the exact renormalization group fixed points equations. The weak disorder sector is characterized by a solution that, for any fixed $N$ larger than 1, is a line of fixed points with Ising thermal exponents and continuously varying magnetic exponents. The number of fixed point solutions allowed by the symmetries of the model increases at strong disorder illustrating the growing dependence on the distributions of the two random couplings. The presence of some critical exponents which do not depend on the symmetry parameter $N$ confirms this type of superuniversality as a peculiar feature of random criticality.
\end{abstract}
\end{titlepage}

\newpage
\tableofcontents

\section{Introduction}
Statistical systems with quenched disorder notoriously represent a particularly difficult problem for the theory of critical phenomena (see e.g. \cite{Cardy_book}). The Harris criterion \cite{Harris} allows to establish whether weak disorder is relevant in the renormalization group (RG) sense and drives the system to a critical point different from that of the pure (i.e. non-disordered) case. Normally, however, the new RG fixed point is not close enough to the pure one to be studied perturbatively, the three-dimensional Ising model providing a remarkable exception \cite{Cardy_book}. The problem is even more severe for the additional fixed points which arise when disorder is further increased. In two dimensions, conformal invariance, which can in principle provide exact results for fixed points occurring at any disorder strength, faced the obstruction coming from the need of combining its usual implementation \cite{BPZ,DfMS} with the subtleties of the replica method \cite{EA} through which random couplings are handled theoretically. As a result of these difficulties, criticality in disordered systems has mainly been investigated numerically.

In the last years, however, it has been shown that the replica method can be exploited in an exact way at RG fixed points in two dimensions implementing conformal invariance in the scattering framework \cite{random}. This opened the way to a series of results (\cite{colloquium} for a review) and unveiled peculiar properties of random criticality such as ``superuniversality", which amounts to a symmetry-independence of some critical exponents. The prediction of \cite{colloquium} that superuniversality could extend to three dimensions has found a first confirmation in \cite{GD_Nishimori}, where exact critical exponents have been determined at the Nishimori multicritical point \cite{Nishimori,Nishimori_book,LdH1}. The scattering framework, initially exploited for magnetic random fixed points \cite{random,colloquium}, has recently been extended to spin glass fixed points \cite{spin_glass}, with the remarkable consequence that exact results in two dimensions allowed to shed light on the fundamental problem of the nature of the spin glass transition in three dimensions \cite{spin_glass_symmetry}. 

In the present paper we use the scattering theory to study magnetic criticality in the two-dimensional disordered $N$-color Ashkin-Teller model, which corresponds to $N$ Ising models coupled by four-spin interactions \cite{AT,GW} in presence of quenched disorder. The pure model is characterized by a phase transition which for $N>2$ is first order due to a mechanism that can be studied perturbatively \cite{GW,GN,Fradkin,Shankar}. However, random couplings should lead to a second order transition to comply with the result of \cite{AW} about the softening of first order transitions by disorder in two dimensions. The investigation of this issue within the scattering framework \cite{random_line} showed that the softening indeed occurs but with a surprising feature: a line of RG fixed points arises for any fixed $N$ other than 1, thus allowing the first exact realization of nonuniversal critical behavior in random criticality. It was also explained in \cite{random_line} how this result relates to perturbative studies \cite{DD_AT,Dotsenko_AT,Murthy,Cardy_GN} and sheds light on numerical results \cite{WD,Katzgraber1,Katzgraber2,Vojta_3color,Vojta_4color}. We now extend our attention to the full space of RG fixed points allowed by the symmetries of the model by determining all the solutions of the exact fixed point equations. The outcome is particularly rich, as expected from the fact that -- at variance with models with a single disorder strength parameter such as Potts or $O(N)$ \cite{colloquium} -- the Ashkin-Teller model allows for two random couplings. Among our findings, the fact that the number of fixed point solutions increases at strong disorder illustrates that universality with respect to the disorder distribution  -- expected at weak disorder on perturbative grounds -- reduces when disorder increases. 

The paper is organized as follows. In the next section we introduce the model and derive the exact fixed point equations. The solutions of these equations are then discussed in section~\ref{solutions_pure} for the pure case and in section~\ref{solutions_disordered} for the disordered case. Section~\ref{superuniversality} is devoted to critical exponents and superuniversality, while section~\ref{conclusion} contains some final remarks.

\section{Exact fixed point equations}
\label{fp_eqs}
The $N$-color Ashkin-Teller model \cite{AT,GW} corresponds to $N$ Ising models coupled by four-spin interaction. We consider the model in two dimensions with lattice Hamiltonian
\EQ
H=-\sum_{\langle x,y\rangle}\left[J_{xy}\sum_{a=1}^N\sigma_a(x)\sigma_a(y)+K_{xy}\sum_{a\neq b}\sigma_a(x)\sigma_a(y)\sigma_b(x)\sigma_b(y)\right]\,,
\label{lattice}
\EN
where $\sigma_a(x)=\pm 1$ is the spin variable of the $a$-th Ising model at site $x$ of a regular lattice, and $\sum_{\langle x,y\rangle}$ denotes the sum over nearest neighbors. Referring to the index $a=1,\dots,N$ which labels the Ising copies as the ``color" index, we see that the Hamiltonian is invariant under spin reversal for each color and permutations among the colors. We consider the model in presence of quenched disorder, which means that $J_{xy}$ and $K_{xy}$ are random couplings drawn from probability distributions $P_1(J_{xy})$ and $P_2(K_{xy})$. We do not further specify $P_1$ and $P_2$ since, as we are going to see, we will generally derive directly in the continuum the exact RG fixed points equations corresponding to the above mentioned symmetries of the Hamiltonian.

In the scaling limit of statistical lattice models, RG fixed points correspond to conformal field theories (CFTs) \cite{Cardy_book}. We then use the scattering framework of \cite{paraf}, in which one of the two spatial dimensions plays the role of imaginary time, and exploit the fact that infinite-dimensional conformal symmetry of two-dimensional RG fixed points \cite{DfMS} yields infinitely many quantities to be conserved in scattering processes. This forces initial and final states to be kinematically identical (complete elasticity), while scale and relativistic invariance lead to scattering amplitudes which do not depend on energy. These drastic simplifications make possible the exact solution of the scattering problem and allowed new results for problems of pure \cite{DT1,ising_vector,DDL_nematic,DLD_RPN,DLD_CPN,potts_qr,RPN_universality} and random criticality \cite{random,colloquium,spin_glass,DT2,DL_ON1,DL_ON2,DL_softening} in two dimensions, and beyond two dimensions \cite{spin_glass_symmetry}.

\begin{figure}[t]
\centering
\includegraphics[width=14cm]{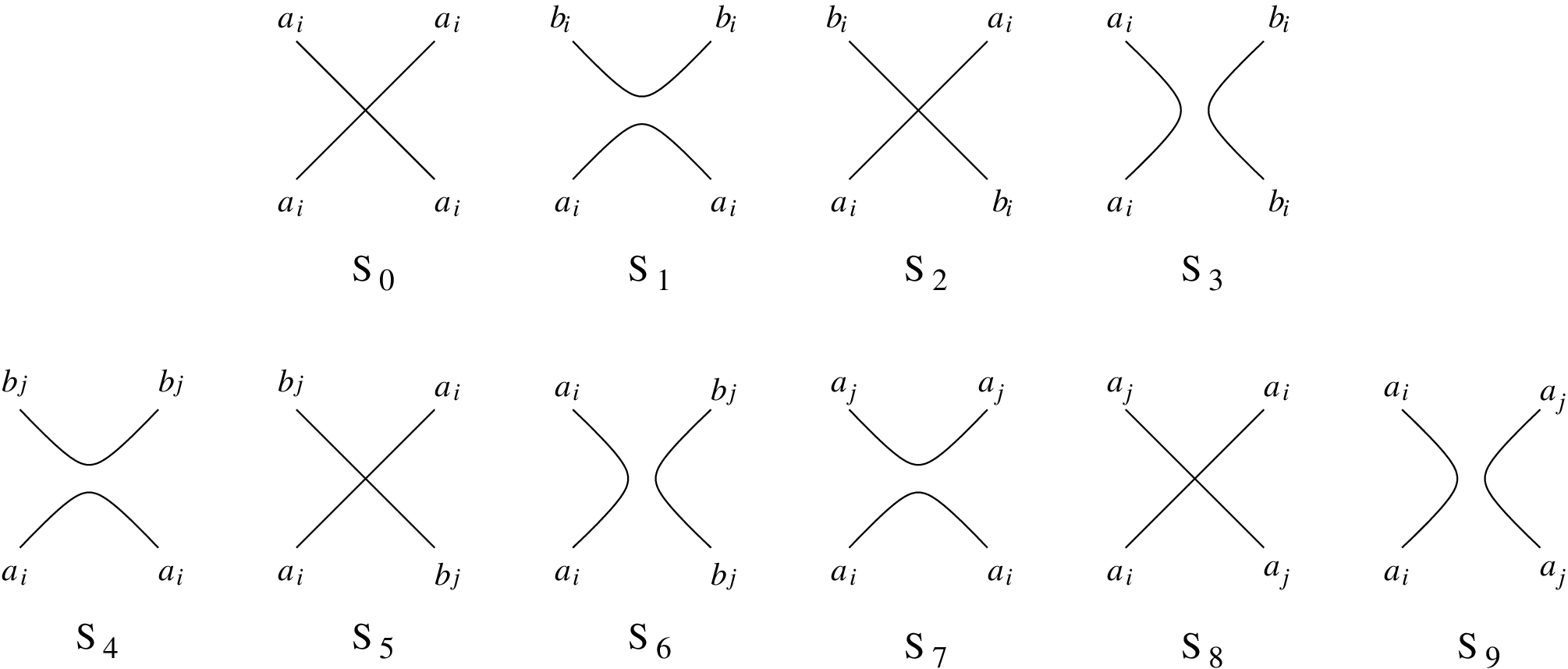}
\caption{Scattering amplitudes for the disordered $N$-color Ashkin-Teller model. $a_i$ labels a particle excitation in the $i$-th replica of the $a$-th Ising model ($a\neq b$, $i\neq j$). Time runs upwards.}
\label{amplitudes}
\end{figure}

The model is specified by the symmetry representation carried by the scattering particles, which are the fundamental collective excitation modes of the system. In the present case, we denote by the color index $a=1,2,\ldots,N$ the particle corresponding to an elementary excitation in the $a$-th Ising model; such a particle is odd under $\sigma_a\to-\sigma_a$. The average over quenched disorder (see e.g. \cite{Cardy_book}) is taken on the free energy $-\ln Z$, where 
\EQ
Z=\sum_{\{\sigma_{1}(x)\},\cdots,\{\sigma_{N}(x)\}}e^{-H/T}
\label{Z}
\EN
is the partition function with an assigned disorder configuration. Then, the identity 
\EQ
\ln Z=\lim_{n\to 0}\frac{Z^n-1}{n}\,
\label{trick}
\EN
shows that the effect of the disorder average is that of coupling $n\to 0$ replicas of the system with Hamiltonian (\ref{lattice}). Within the scattering framework this corresponds to introducing a replica index $i=1,2,\ldots,n$, so that $a_i$ labels a particle in the $i$-th replica of the $a$-th Ising copy. Figure~\ref{amplitudes} then shows the elastic scattering amplitudes allowed by spin reversal symmetry in each of the $Nn$ Ising models. In the first row we have the amplitudes which involve a single replica and are the only ones defined in the ``pure" model, namely in the model without disorder ($n=1$). The amplitudes coupling different replicas arise in the disordered case and are shown in the second row. 

Denoting by $S_{\alpha\beta}^{\gamma\delta}={}_\alpha^\delta\times_\beta^\gamma$ a generic amplitude, the particularly simple form of the crossing and unitarity equations \cite{colloquium,paraf}
\begin{equation}
S_{\alpha\beta}^{\gamma\delta}=[S_{\alpha\delta}^{\gamma\beta}]^*\,,
\label{cross}
\end{equation} 
\begin{equation}
\sum_{\epsilon,\phi} S_{\alpha\beta}^{\epsilon\phi}[S_{\epsilon\phi}^{\gamma\delta}]^*=\delta_{\alpha\gamma}\delta_{\beta\delta}\,,
\label{unitarity}
\end{equation}
respectively, is allowed by the energy-independence of the amplitudes. The latter are also invariant under spatial reflection ($S_{\alpha\beta}^{\gamma\delta}=S_{\beta\alpha}^{\delta\gamma}$) and time reversal ($S_{\alpha\beta}^{\gamma\delta}=S^{\alpha\beta}_{\gamma\delta}$).
For the amplitudes of Figure~\ref{amplitudes} the crossing equations (\ref{cross}) take the form
\begin{align}
S_k&=S_k^*\,,\hspace{3.4cm}k=0,2,5,8
\label{crossing1}\\
S_k&=S_{k+2}^*\equiv X_k+iY_k\,,\hspace{1cm}k=1,4,7\,,
\label{crossing2}
\end{align}
where we introduced $X_k$ and $Y_k$ real. Then the unitarity equations (\ref{unitarity}) translate into
\begin{align}
&S_0^2+(N-1)(X_1^2+Y_1^2)+(N-1)(n-1)(X_4^2+Y_4^2)+(n-1)(X_7^2+Y_7^2)=1\,,
\label{uni1} \\
&2S_0 X_1+(N-2)(X_1^2+Y_1^2)+2(n-1)(X_4 X_7+Y_4 Y_7)+(N-2)(n-1)(X_4^2+Y_4^2)=0\,,\\
&X_1 S_2=0\,,\label{uni3}\\
&X_1^2+Y_1^2+S_2^2=1\,,
\label{uni4}\\
&2 S_0 X_4+2(X_1X_7+Y_1Y_7)+2(N-2)(X_1X_4+Y_1Y_4)+2(n-2)(X_4X_7+Y_4Y_7) \nonumber\\
&+(N-2)(n-2)(X_4^2+Y_4^2)=0\,,\\
&X_4 S_5=0\,,\label{uni6}\\
&X_4^2+Y_4^2+S_5^2=1\,,\label{uni7}\\
&2 S_0X_7+(n-2)(X_7^2+Y_7^2)+2(N-1)(X_1X_4+Y_1Y_4)+(N-1)(n-2)(X_4^2+Y_4^2)=0,\\
&X_7 S_8=0\,, \label{uni9}\\
&X_7^2+Y_7^2+S_8^2=1\,.
\label{uni10}
\end{align}

This derivation implies that the solutions of Eqs.~(\ref{uni1})-(\ref{uni10}) are the RG fixed points of the replicated system of $N$ Ising models with couplings preserving global invariance under spin reversals $\sigma_{a,i}\to -\sigma_{a,i}$ and permutations $\sigma_{a,i}\leftrightarrow\sigma_{b,j}$. The equations contain $N$ and $n$ as parameters that do not need to be integers, so that, in particular, the limit $n\to 0$ corresponding to quenched disorder can be taken straightforwardly.

Notice also that the present theory reduces to the replicated $O(N)$ theory \cite{colloquium,DL_ON1,DL_ON2} when the scattering processes of Figure~\ref{amplitudes} do not distinguish $a\neq b$ from $a=b$. As a consequence, the $O(N)$-invariant case corresponds to
\EQ
O(N):\hspace{.5cm}S_0=S_1+S_2+S_3\,,\hspace{1cm}S_4=S_7\,,\hspace{1cm}S_5=S_8\,,\hspace{1cm}S_6=S_9\,.
\label{ON}
\EN

We also observe that the state $\sum_{a,i}|a_ia_i\rangle$ scatters into itself producing the phase
\EQ 
S= S_0 + (N-1)[S_1 +(n-1)S_4] + (n-1)S_7\,.
\label{S}
\EN 
It was shown in \cite{colloquium,paraf} that such an amplitude corresponding to the symmetry invariant scattering channel can be written as
\EQ
S=e^{-2i\pi\Delta_\eta}\,,
\label{phase}
\EN 
where $\Delta_\eta$ is the conformal dimension of the chiral field $\eta$ that creates a right-moving particle. We recall that the conformal dimensions $\Delta_\Phi$ and $\bar{\Delta}_\Phi$ of a field $\Phi(x)$ determine the scaling dimension $X_\Phi=\Delta_\Phi+\bar{\Delta}_\Phi$, and that a field is chiral if one of its conformal dimensions vanishes.

It must also be pointed out that the form of the crossing and unitarity equations (\ref{cross}) and (\ref{unitarity}) is such that, given a solution, another solution is obtained reversing the sign of all the scattering amplitudes. As a consequence, the space of solutions of the
exact fixed point equations provided by scattering theory is made of pairs of solutions related by such “sign reversal”. It is then possible that only one member of a pair is physically interesting for the model under consideration, the other member being automatically generated by the form of the equations.

\section{Solutions for the pure model}
\label{solutions_pure}
The pure model corresponds to $J_{xy}=J$ and $K_{xy}=K$ in the Hamiltonian (\ref{lattice}). In absence of disorder there is no need for replicas and it is sufficient to consider $n=1$. In this case, Eqs.~(\ref{uni1})-(\ref{uni4}) receive contributions only from the single-replica amplitudes $S_0,..,S_3$ and decouple from the remaining equations. It follows that (\ref{uni1})-(\ref{uni4}) with $n=1$ are the fixed point equations for the pure model, and their solutions are listed in Table~\ref{pure_solutions}. 

The solutions P1 with $S_2=S_0$, P2 and P3 satisfy $S_0=S_1+S_2+S_3$ and are the fixed points of the $O(N)$-invariant subspace (recall (\ref{ON})). We now recall some main features of these solutions, referring the reader to \cite{colloquium,DL_ON2} for the details. P1$_{\pm}$ is the free boson/fermion solution defined for any $N$. P2 is defined for $N\in[-2,2]$ and corresponds to the gas of critical self-avoiding loops \cite{Nienhuis}. P3 exists only for $N=2$ and is the line of fixed points parametrized by $Y_1$ accounting for the BKT transition of the XY model \cite{Berezinskii,KT} and for the continuously varying exponents of the Ashkin-Teller model \cite{Baxter,KB}. Indeed both models renormalize at criticality onto the Gaussian field theory with Hamiltonian \cite{DfMS}
\EQ
{\cal H}_{\text{Gauss}}=\frac{1}{4\pi}\int d^2x\,(\nabla\varphi)^2\,,
\label{gauss}
\EN
and energy density field $\varepsilon(x)=\cos 2b\varphi(x)$ with scaling dimension $X_\varepsilon=2b^2$; $b^2$ is the coordinate along the line of fixed points. Using complex coordinates $x_\pm=x_1\pm ix_2$, the equation of motion $\partial_+\partial_-\varphi=0$ yields the decomposition $\varphi(x)=\phi_+(x_+)+\phi_-(x_-)$. The fields 
\EQ
U_m(x)=e^{i\frac{m}{2b}[\phi_+(x_+)-\phi_-(x_-)]}\,,\hspace{1cm}m\in \mathbb{Z}\,
\label{vertex}
\EN
have scaling dimension $m^2/8b^2$ and satisfy the condition that $\langle\cdots\varepsilon(x)U_m(0)\cdots\rangle$ is single valued in $x$ (see e.g. \cite{colloquium}). $(\phi_+-\phi_-)/2b$ is the $O(2)$ angular variable, the order vector field $(s_1,s_2)$ of the XY model corresponds to $s_1\pm is_2=U_{\pm 1}$, and the BKT transition point corresponds to $b^2=1$, where $\varepsilon$ becomes marginal. On the other hand, the order fields of the $N=2$ Ashkin-Teller model -- which are the Ising spin fields $\sigma_1$ and $\sigma_2$ -- cannot be expressed in terms of the bosonic fields $\phi_\pm$ \cite{KB,DG_at}, and the symmetry relevant for this model is discrete. Such a discrete symmetry -- at variance with the continuous $O(2)$ symmetry of the XY model \cite{MW,Hohenberg,Coleman} -- breaks spontaneously also in two dimensions and yields an ordered phase at low temperatures \cite{Baxter}. In this case, the parameter $b^2$ of the Gaussian theory (\ref{gauss}) spans the line of fixed points in the $J$-$K$ plane of the couplings of the lattice Hamiltonian (\ref{lattice}) \cite{KB,DG_at}. $Y1=-\sin(\pi/2b^2)$ relates the Gaussian theory to the scattering solution P3 \cite{paraf}. 

\begin{table}
\begin{center}
\begin{tabular}{c|c|c|c|c|c}
\hline 
Solution & $N$ & $S_0$ & $S_2$ & $X_1$ & $Y_1$\\ 
\hline \hline
$\text{P}1_{\pm}$ & $\mathbb{R}$ & $\pm 1$ & $(\pm) 1$ & $0$ & $0$ \\ 
$\text{P}2_{\pm}$ & $[-2, 2]$ & $\pm\sqrt{2-N}$ & $0$ & $\pm\frac{1}{2}\sqrt{2-N}$ & $(\pm)\frac{1}{2}\sqrt{2+N}$ \\ 
$\text{P}3_{\pm}$ & $2$ & $\pm\sqrt{1-Y_1^2}$ & $\pm\sqrt{1-Y_1^2}$ & $0$ & $[-1,1]$ \\
$\text{P}4_{\pm}$ & $2$ & $\pm\sqrt{1-Y_1^2}$ & $\mp\sqrt{1-Y_1^2}$ & $0$ & $[-1,1]$ \\
$\text{P}5_{\pm}$ & $2$ & $0$ & $0$ & $\pm\sqrt{1-Y_1^2}$ & $[-1,1]$ \\
[0.7em] 
\hline 
\end{tabular} 
\caption{Solutions of Eqs.~(\ref{uni1})-(\ref{uni4}) with $n=1$. They correspond to the RG fixed points of the pure $N$-color Ashkin-Teller model. Signs in brackets are both allowed.
}
\label{pure_solutions}
\end{center}
\end{table}

Consider now the field theory with Hamiltonian
\EQ
{\cal H}_\text{Gauss}+\int d^2x\,\{g\,\varepsilon(x)+\tilde{g}\,[U_4(x)+U_{-4}(x)]\}\,.
\label{Z4}
\EN
Since we have seen that $U_{\pm 1}$ define the components of an $O(2)$ vector, the terms $U_{\pm 4}$ in (\ref{Z4}) break the $O(2)$ symmetry of ${\cal H}_\text{Gauss}$ down to $\mathbb{Z}_4$. The $N=2$ Ashkin-Teller model does possess $\mathbb{Z}_4$ symmetry, since the Hamiltonian (\ref{lattice}) is invariant under rotations of the vector $(\sigma_1,\sigma_2)$ by angles multiples of $\pi/2$. It also follows from the scaling dimensions that we specified above that at $b^2=1$ all the fields in the integral in (\ref{Z4}) are marginal. The perturbative study of the RG equations around $b^2-1=g=\tilde{g}=0$ gives at leading order three lines of fixed points: $g=\tilde{g}=0$ and $b^2=1$, $g=\pm\tilde{g}$ \cite{JKKN}. It was then conjectured in \cite{Kadanoff} that these lines may persist to all orders. The exact results of Table~\ref{pure_solutions} show that this is indeed the case since, besides P3, we also have the lines of fixed points P4 and P5. The three lines of fixed points meet at the point $b^2-1=g=\tilde{g}=0$ (i.e. $Y_1=-1$). The maximal value of $b^2$ realized in the square lattice Ashkin-Teller model is $3/4$ \cite{KB}, so that only P3 is observed in the phase diagram\footnote{See \cite{DG_at,D_AT} for the off-critical correlation functions in the scaling limit towards this critical line.}.

\section{Solutions for the disordered model}
\label{solutions_disordered}
\subsection{Generalities}
The RG fixed points for the model (\ref{lattice}) with quenched disorder correspond to the solutions of the exact fixed point equations (\ref{uni1})-(\ref{uni10}) with $n=0$ and coupled replicas. It follows from Figure~\ref{amplitudes} and equations (\ref{crossing2}), (\ref{uni7}) and (\ref{uni10}) that coupled replicas amount to $S_4$ and/or $S_7$ nonvanishing\footnote{For $S_4=S_7=0$ the pure solutions of Table~\ref{pure_solutions} are recovered.}. $|S_7|$ is a measure of the interaction among replicas with the same color, and then a measure of the strength of disorder associated to the coupling $J_{xy}$ in (\ref{lattice}). On the other hand, $|S_4|$ is a measure of the interaction among replicas with different color, and then a measure of the strength of disorder associated to the coupling $K_{xy}$. 

\begin{figure}[htbp]
    \centering
    
    \begin{subfigure}{\textwidth}
        \centering
        \includegraphics[width=0.85\linewidth]{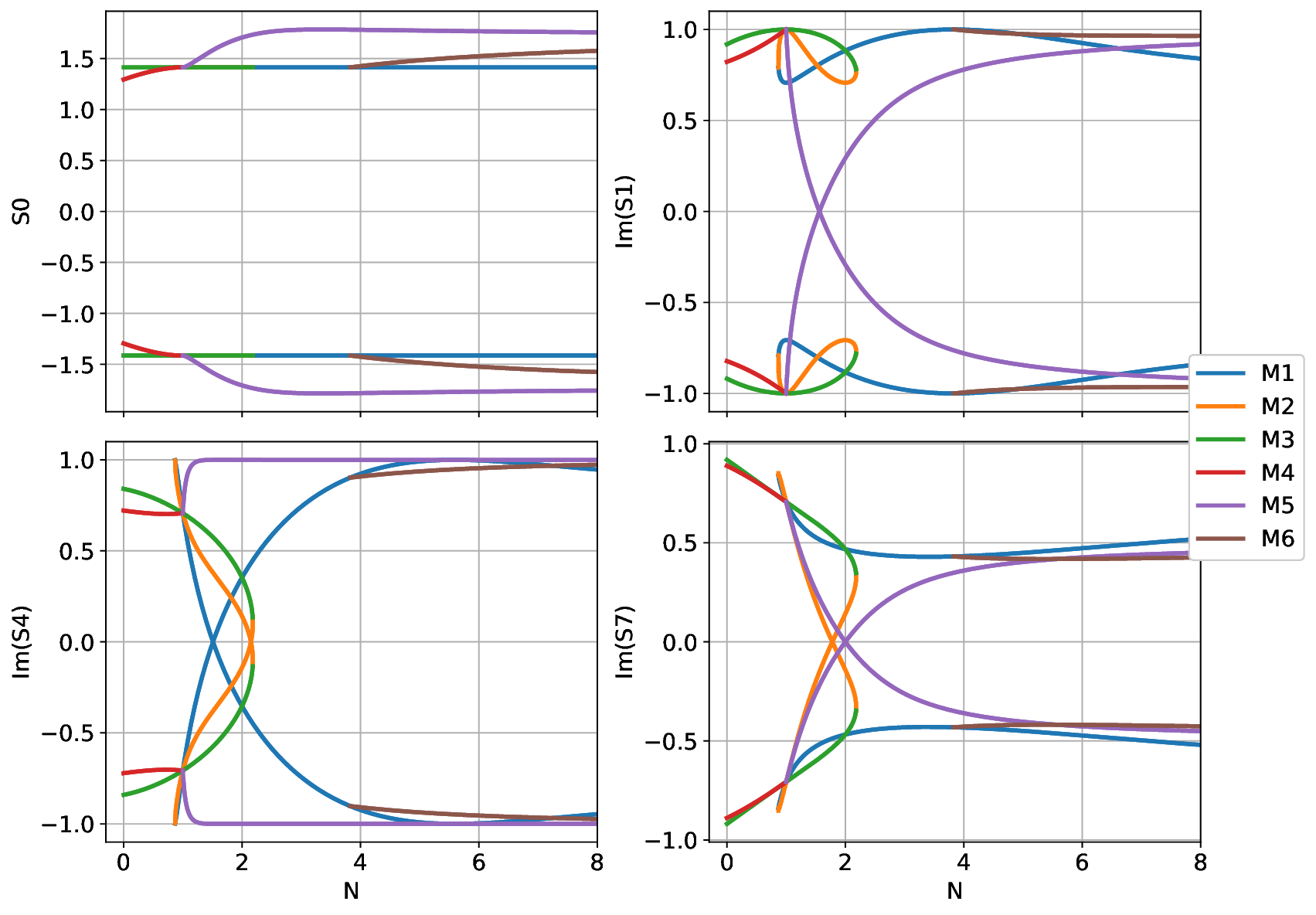}
        \caption{$ $}
        \label{fig:groupM}
    \end{subfigure}
    
    \vspace{0.5cm} 
    
    \begin{subfigure}{\textwidth}
        \centering
        \includegraphics[width=0.85\linewidth]{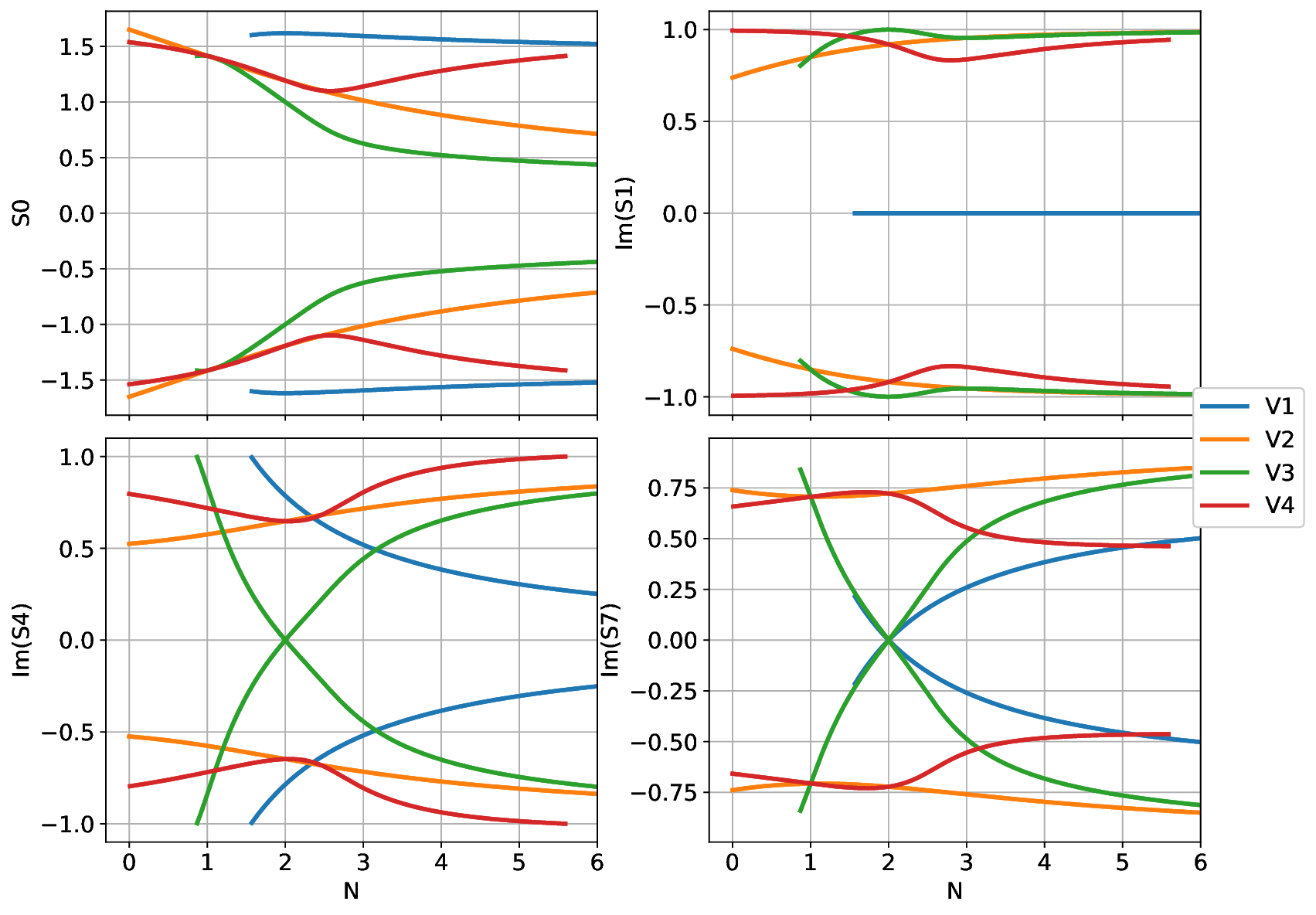}
        \caption{$ $}
        \label{fig:groupS}
    \end{subfigure}

    \caption{Amplitudes for the solutions not given analytically in Table 2 (with the exception of V1).}
    \label{fig:ampiezze_groups}
\end{figure}

We list in Table~\ref{dis_sol} the solutions of equations (\ref{uni1})-(\ref{uni10}) with $n=0$ and coupled replicas. We only list the solutions whose range of definition in $N$ includes at least one positive integer. Table~\ref{dis_sol} contains the amplitudes $S_0$, $S_1$, $S_4$ and $S_7$, the remaining amplitudes following straightforwardly from equations (\ref{crossing2}), (\ref{uni4}), (\ref{uni7}) and (\ref{uni10}). The functions $f_\pm$ and $g$ appearing in Table~\ref{dis_sol} are given by 
\begin{align}
        &f_{\pm}(N)=\sqrt{ \frac{2N^2-3N\pm 2\sqrt{ - 2N^3+6N^2 - 4N +1 }}{(2N-1)^2N}}\,,\label{f}\\
        &g(N) = \sqrt{ \frac{N^2 +2N -2 + \sqrt{4+N(N^3+4N-8)}}{N^2} }\,.\label{g}
\end{align}
For some of the solutions of type M and type V the complete analytical expression is either too difficult to be determined or too cumbersome. In these cases the amplitudes are more conveniently obtained solving the equations numerically and some of them are plotted in Figure~\ref{fig:ampiezze_groups}. 

Besides the invariance of the fixed point equations under the sign reversal of all amplitudes pointed out at the end of section~\ref{fp_eqs}, the sign doublings exhibited by the solutions of Table~\ref{dis_sol} also reflect other symmetries of Eqs.~(\ref{uni1})-(\ref{uni10}), namely their invariance under the simultaneous change of sign of $S_0$, $X_1$, $X_4$ and $X_7$, or the simultaneous change of sign of $Y_1$, $Y_4$ and $Y_7$. 

We now discuss the solutions depending on their properties with respect to the strength of disorder.

\begin{table}[t]
\centering
\begin{adjustbox}{max width=1\textwidth}
\renewcommand{\arraystretch}{1.45}
\begin{tabular}{c|c|c|c|c|c}
\hline
Solution & $N$ & $S_0$ & $S_1$ & $S_4$ & $S_7$ \\
\hline\hline

L1$_\pm$ & $\mathbb{R}$ &
$\pm 1$ &
$i[-1,1]$ &
$S_1$ &
0 \\

L2$_\pm$ & 2 &
$\pm 1 [\pm] \frac{Y_4}{\sqrt{1+Y_4^2}}$ &
$i\left(Y_4 [\mp] \text{sgn}(S_7) \frac{1}{\sqrt{1+Y_4^2}}\right)$ &
$i[-1,1]$ &
$\pm 1$ \\

\hline


W1$_\pm$  & $[2,\infty)$ &
$\pm\frac{N-2}{N}$ &
$\mp\frac{2}{N}[\pm]i\frac{\sqrt{N^2-4}}{N}$ &
$[\pm]i\frac{\sqrt{N^2-4}}{N}$ &
0 \\

W2$_\pm$ & $[2,\infty)$ &
0 &
$\pm\frac{1}{N-1} [\pm] i\frac{\sqrt{N^2-2N}}{N-1}$ &
$[\pm] i\sqrt{\frac{N-2}{N}}$ &
$[\pm] i\sqrt{\frac{N-2}{N}}$ \\

W3$_\pm ^+$    & $[1,2.19..]$ &
$\pm\sqrt{2-N+N(N-1)f_+^2(N)}$ &
$\mp\sqrt{1-N^2f_+^2(N)}[\pm] i Nf_+(N)$ &
$[\pm] i f_+(N)$ &
$[\mp] i(N-1)f_+(N)$ \\

W3$^-_\pm$   & $[2,2.19..]$ &
$\pm\sqrt{2-N+N(N-1)f_-^2(N)}$ &
$\mp\sqrt{1-N^2f_-^2(N)}[\pm] i Nf_-(N)$ &
$[\pm] i f_-(N)$ &
$[\mp] i(N-1)f_-(N)$ \\

W4$_\pm$ & $[\sqrt{2}-1,\infty)$ &
$\pm\frac{2}{N+1}$ &
$\pm\frac{1}{N+1}[\pm]i\sqrt{\frac{N(N+2)}{(N+1)^2}}$ &
$[\pm] i\frac{N-1}{N+1}\sqrt{\frac{N+2}{N}}$ &
$[\pm] i\frac{N-1}{N+1}\sqrt{\frac{N+2}{N}}$ \\

\hline


V1$_\pm$   & $[\frac{\sqrt{17}-1}{2},\infty)$ &$\pm g(N)$& 
0 & 
$[\pm] i   \sqrt{\frac{g(N)^2-2}{N-1}}$&
$\pm\sqrt{1 - \frac{(2-N)^2}{4}Y_4^2} [\pm] i\frac{2-N}{2}Y_4$\\

V2$_\pm$  & $[-2.4142..,\infty)$ & & $|S_1|=1$ & & $|S_7|=1$ \\

V3$_\pm$  & $[0.86730..,\infty)$ & & $|S_1|=1$ & & $|S_7|=1$ \\

V4$_\pm$  & $(-\infty,5.5986..]$ &  & $|S_1|=1$ & & $|S_7|=1$ \\

V5$_\pm$ & $\mathbb{R}$ &
$\pm\sqrt{2}$ & 0 & 0 &
$\pm\frac{1}{\sqrt{2}}(\pm)i\frac{1}{\sqrt{2}}$ \\

\hline


M1$_\pm$  & $[0.86525..,\infty)$ &
$\pm\sqrt{2}$ &
$|S_1|=1$ &
$|S_4|=1$ &
$|S_7|=1$ \\

M2$_\pm$  & $[0.86525..,2.1830..]$ &
$\pm\sqrt{2}$ &
$|S_1|=1$ &
$|S_4|=1$ &
$|S_7|=1$ \\

M3$_\pm$  & $(-\infty,2.1830..]$ &
$\pm\sqrt{2}$ &
$|S_1|=1$ &
$|S_4|=1$ &
$|S_7|=1$ \\

M4$_\pm$  & $[-2,1]$ & & & $|S_4|=1$ & $|S_7|=1$ \\

M5$_\pm$  & $[1,\infty)$ & & & $|S_4|=1$ & $|S_7|=1$ \\

M6$_\pm$  & $[3.8284..,\infty)$ & & &$|S_4|=1$ & $|S_7|=1$ \\

M7$_\pm$ & $\mathbb{R}$ &
$\pm \sqrt{2}$ &
$\pm \frac{1}{\sqrt{2}}[\pm]\,i\frac{1}{\sqrt{2}}$ &
$\pm \frac{1}{\sqrt{2}}[\pm]\,i\frac{1}{\sqrt{2}}$ &
$\pm \frac{1}{\sqrt{2}}[\pm]\,i\frac{1}{\sqrt{2}}$ \\

M8$_\pm$ & $\mathbb{R}$ &
$\pm \sqrt{2}$ &
$\mp \frac{1}{\sqrt{2}}[\pm]\,i\frac{1}{\sqrt{2}}$ &
$\mp \frac{1}{\sqrt{2}}[\pm]\,i\frac{1}{\sqrt{2}}$ &
$\pm \frac{1}{\sqrt{2}}[\mp]\,i\frac{1}{\sqrt{2}}$ \\

M9$_\pm$ & $\mathbb{R}$ &
$\pm\sqrt{2}$ &
$\pm \frac{1}{\sqrt{2}}[\pm]i\frac{1}{\sqrt{2}}$ &
$\pm \frac{N^2+2N-1}{\sqrt{2}(N^2+1)}[\pm]i\frac{N^2-2N-1}{\sqrt{2}(N^2+1)}$ &
$\pm \frac{N^2+2N-1}{\sqrt{2}(N^2+1)}[\pm]i\frac{N^2-2N-1}{\sqrt{2}(N^2+1)}$ \\
\hline
\end{tabular}
\end{adjustbox}
\caption{Solutions of Eqs.~(\ref{uni1})-(\ref{uni10}) with $n=0$ and coupled replicas ($S_4$ and $S_7$ do not both vanish). They correspond to the RG fixed points of the $N$-color Ashkin-Teller model in presence of disorder. Only the solutions defined for at least one positive integer value of $N$ are considered. The functions $f_\pm$ and $g$ are given by (\ref{f}) and (\ref{g}). For the amplitudes that we determine numerically we leave an empty entry or specify the modulus when this is $N$-independent. Signs in brackets are both allowed and for those in square brackets the same choice (upper or lower) must be made for the different amplitudes within a solution.}
\label{dis_sol}
\end{table}

\subsection{Solutions with weak disorder limit}
\label{weak_disorder}
The case of weak disorder is normally the best understood theoretically because of the guidance provided by the Harris criterion \cite{Harris} and of the possibility, in some cases, to perform perturbative calculations starting from the pure case. For generic $N$, the pure system admits only the free fixed point solutions P1 of Table~\ref{pure_solutions}. The two-dimensional Ising pure ferromagnet corresponds to a free neutral fermion (see \cite{DfMS}), so that P1$_-$ with $S_0=S_2$ is the fixed point for the system of $N$ Ising pure ferromagnets. In the replicated case, the solution
\EQ
S_0=S_2=S_5=S_8=-1\,,\hspace{1.5cm}S_1=S_3=S_4=S_6=S_7=S_9=0
\label{ff}
\EN
is the fixed point for the system of $Nn$ Ising pure ferromagnets. Then Eqs. (\ref{uni3}), (\ref{uni6}) and (\ref{uni9}) imply that a fixed point solution admitting a limit towards (\ref{ff}) must have $X_1=X_4=X_7=0$. As observed in \cite{random_line}, for all positive values of $N$, there is a single solution of Eqs. (\ref{uni1})-(\ref{uni10}) with $n=0$ which satisfies these requirements, and it is solution L1$_-$ of Table~\ref{dis_sol}. Since $S_4$ is free to vary in an interval, this solution yields a {\it line of random fixed points} for $N$ fixed\footnote{The case $N=1$ is excluded, since $S_4$ is not defined for a single color.}. The purely ferromagnetic solution (\ref{ff}) is recovered in the limit $S_4\to 0$. On the other hand, L1$_-$ extends to strong disorder when $|S_4|$ is not small.

\begin{figure}[t]
\centering
\includegraphics[width=15cm]{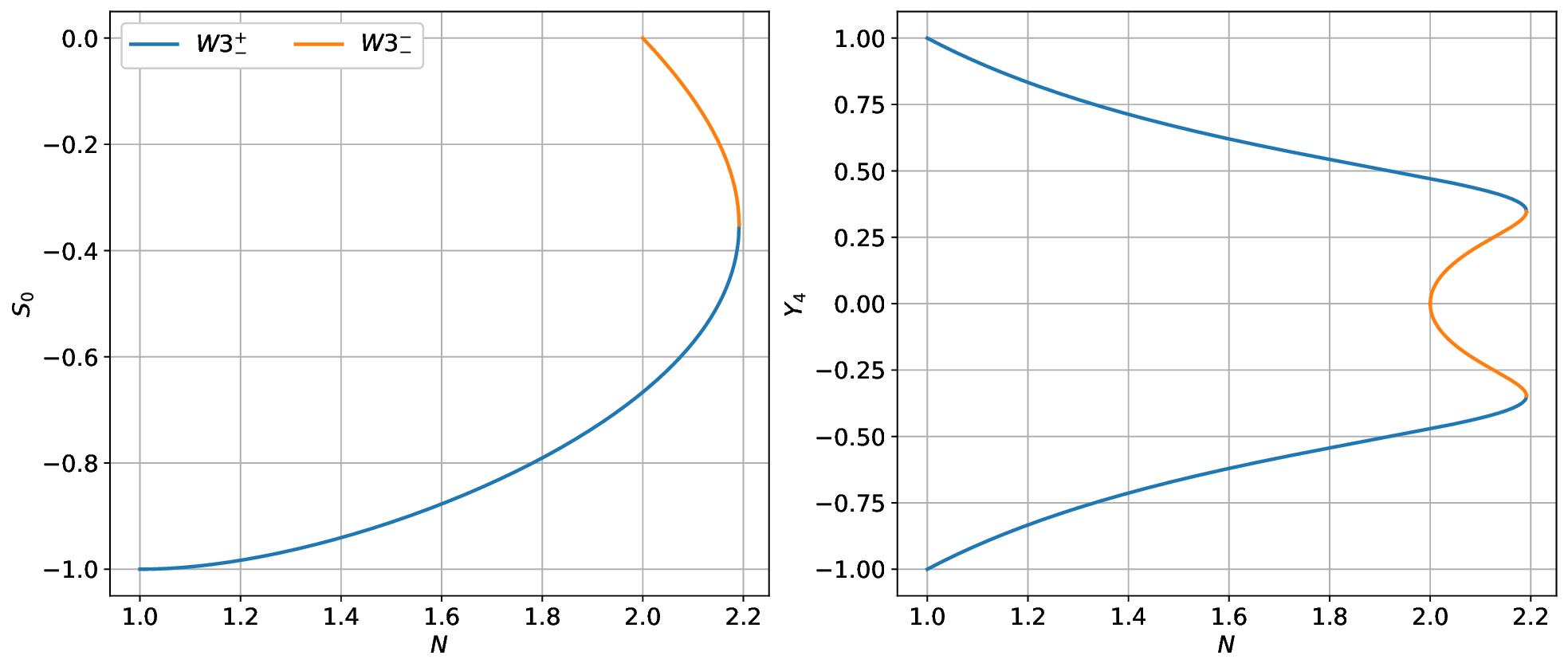}
\caption{Solutions W3$^+$ and W3$^-$ can be seen as continuation of each other.}
\label{W3}
\end{figure}

While L1 is the only solution with a limit of vanishing disorder (i.e. $S_4=S_7=0$) for any $N$, Table~\ref{dis_sol} also contains solutions with vanishing disorder at isolated values of $N$. These are W3$^+$ and W4, for which disorder vanishes at\footnote{Vanishing disorder at $N=1$ only requires $S_7=0$, since $S_4$ is not defined there.} $N=1$, and W1, W2 and W3$^-$, for which disorder vanishes at $N=2$. In order to get additional insight we observe that, in a neighborhood of a value $N=N_0$ at which disorder vanishes within a disordered solution, the replicas are weakly coupled, so that the solution describes a random fixed point which is sufficiently close to the pure fixed point to be perturbatively approachable. Within the perturbative RG (see e.g. \cite{Cardy_book}) this notion of closeness amounts to the fact that the field which drives the flow between the pure and the random fixed point is almost marginal, and is marginal when the two fixed points coalesce, namely at $N_0$. The Harris criterion implies that the scaling dimension of the field which drives the flow is twice the dimension $X_\varepsilon$ of the energy density field of the pure model, and in two dimensions a marginal field has scaling dimension 2. It follows that $X_\varepsilon=1$ at $N_0$, and for two-dimensional systems without disorder the condition $X_\varepsilon=1$ is characteristic of free fermions. Hence, we conclude that the limit $N\to N_0$ in which disorder vanishes has to be a free fermion limit. And indeed, we see that at $N_0=1$ (where only the single-color amplitudes $S_0$, $S_7$, $S_8$ and $S_9$ are physical) W3$^+$ and W4 allow the free fermionic values $S_0=S_8=-1$. On the other hand, at $N_0=2$, W1, W2 and W3$^-$ coincide with the point $Y_1=0$ of the pure solution P5. P5$_-$ with $Y_1=0$ is a free fermion point, as can be seen observing that the diagonalization of the scattering yields for the phase (\ref{S}) the value $S=S_0+(N-1)S_1=-1$ characteristic of free fermions. Solution W4 belongs to the $O(N)$-invariant subspace (recall (\ref{ON})) and its role within the disordered $O(N)$ model was discussed in \cite{colloquium,DL_ON1,DL_ON2}. Solutions W3$^+$ and W3$^-$ meet at $N=1+6^{-2/3}[(9+\sqrt{33})^{1/3}+(9-\sqrt{33})^{1/3}]=2.19..$ and can be seen as the continuation of each other (Figure~\ref{W3}). The role of W1 and W2 emerges once we consider the general properties of the model at weak disorder, to which we now turn.

The description in the continuum of the Hamiltonian (\ref{lattice}) for the case of pure ferromagnet ($J_{xy}=J>0$, $K_{xy}=K$) corresponds to
\EQ
{\cal H}_\textrm{pure}=\sum_{a=1}^N\left[{\cal H}_a^0-\tau\int d^2x\,\varepsilon_a(x)\right]-\lambda\sum_{a\neq b}\int d^2x\,\varepsilon_a(x)\varepsilon_b(x)\,,
\label{pure}
\EN
where ${\cal H}_a^0$ and $\varepsilon_a(x)$ are, respectively, the fixed point Hamiltonian (free massless neutral fermion) and the energy density field (product of the fermion components $\psi_a$ and $\bar{\psi}_a$) of the $a$-th Ising model. The introduction of random couplings $J_{xy}$ and $K_{xy}$ in (\ref{lattice}) amounts to random couplings $\tau(x)$ and $\lambda(x)$ in (\ref{pure}). The identity (\ref{trick}) then leads to the introduction of replicas $i=1,\ldots,n$ and to the Hamiltonian
\bea
{\cal H}&=&\sum_{a=1}^N\sum_{i=1}^n\left[{\cal H}_{a,i}^0-\tau\int d^2x\,\varepsilon_{a,i}(x)\right]-\int d^2x\left[\lambda_1\sum_{a}\sum_{i\neq j}\varepsilon_{a,i}(x)\varepsilon_{a,j}(x)\right.\nonumber\\
&+&\left.\lambda_2\sum_{a\neq b}\sum_{i}\varepsilon_{a,i}(x)\varepsilon_{b,i}(x)+\lambda_3\sum_{a\neq b}\sum_{i\neq j}\varepsilon_{a,i}(x)\varepsilon_{b,j}(x)\right].
\label{cumulant}
\eea
Since the scaling dimension of $\varepsilon_{a,i}$ at the Ising ferromagnetic fixed point around which we are expanding is $X_{\varepsilon}=1$, the terms quadratic in the energy densities appearing in (\ref{cumulant}) are marginal in the RG sense, while the terms more than quadratic were omitted because irrelevant\footnote{We also recall that in the two-dimensional Ising model without disorder the operator product expansion (OPE) $\varepsilon_{a,i}\cdot\varepsilon_{a,i}$ produces the identity plus irrelevant fields (see \cite{DfMS}).}. Calling $A_\alpha(x)$ ($\alpha=1,2,3$) the sum of marginal fields conjugated to $\lambda_\alpha$ in (\ref{cumulant}), the general form of the one-loop RG equations \cite{Cardy_book} yields
\EQ
\frac{d\lambda_\alpha}{dt}=\sum_{\beta,\gamma}C_{\beta,\gamma}^\alpha\,\lambda_\beta\lambda_\gamma+O(\lambda^3)\,,
\label{RG}
\EN
where we denote by $t$ the logarithmic RG scale and by $C_{\beta,\gamma}^\alpha$ the coefficient of $A_\alpha$ in the OPE $A_\beta\cdot A_\gamma$. The combinatorial method (see \cite{Cardy_book}) allows to easily determine these OPE coefficients observing that, since $A_\alpha$ is a sum of products of two energy densities, $A_\beta\cdot A_\gamma$ produces $A_\alpha$ upon a single contraction $\varepsilon_{a,i}\cdot\varepsilon_{b,j}\to\delta_{ab}\delta_{ij}$. The result is \cite{random_line}
\bea
\frac{d\lambda_1}{dt}&=&4 (n - 2)\lambda_1^2 + 4 (N - 1) (n - 2) \lambda_3^2 + 8 (N - 1) \lambda_2 \lambda_3\,,\label{RG1}\\
\frac{d\lambda_2}{dt}&=& 4 (N - 2)\lambda_2^2 + 4 (N - 2) (n - 1) \lambda_3^2 + 8 (n - 1) \lambda_1 \lambda_3\,,\\
\frac{d\lambda_3}{dt}&=& 4 (N - 2) (n - 2)\lambda_3^2 + 8 \lambda_1 \lambda_2 + 8 (n - 2) \lambda_1 \lambda_3 + 8 (N - 2) \lambda_2 \lambda_3\,.\label{RG3}
\eea

 For the pure case $\lambda_1=\lambda_3=0$ (decoupled replicas) the surviving equation $d\lambda_2/dt=4(N-2) \lambda_2^2$ correctly reproduces that of the Gross-Neveu model \cite{GN} ($N$ neutral fermions coupled by $O(N)$-symmetric four-fermion interaction). It follows from this equation that $\lambda_2>0$ is marginally relevant (resp. irrelevant) for $N>2$ (resp. $N<2$). For $N=2$, we have already seen non-perturbatively that marginality persists at all orders and produces a line of fixed points. 

The energy density field $\varepsilon$ is odd under the Kramers-Wannier duality \cite{KW} which leaves invariant the transition point of the pure two-dimensional Ising ferromagnet. Since the transition of the pure model (\ref{pure}) is still expected to occur at a self-dual point and $\varepsilon_a\varepsilon_b$ is even under duality, a nonzero value of $\lambda=\lambda_2$ is allowed at the transition $\tau=0$. For $N>2$, the marginal relevance of $\lambda_2>0$ then implies that a finite correlation length is developed and the transition is first order \cite{GW,GN,Fradkin,Shankar}. 

\begin{figure}[t]
    \centering
    \begin{subfigure}[h]{0.49\textwidth}
        \includegraphics[width=\textwidth]{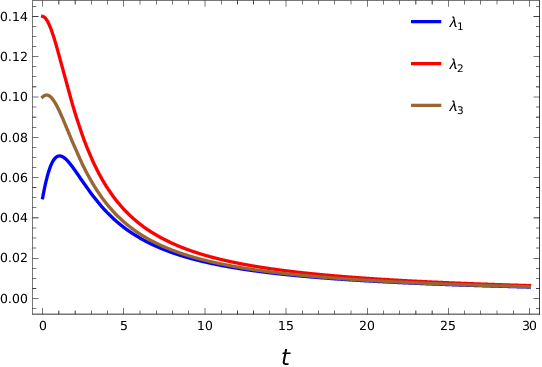}
    \end{subfigure}\hspace{.3cm}%
    \begin{subfigure}[h]{0.49\textwidth}
        \includegraphics[width=\textwidth]{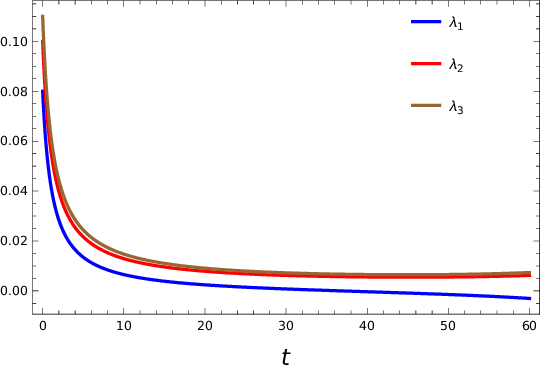}
    \end{subfigure}
    \caption{RG flows determined by Eqs.~(\ref{RG1})-(\ref{RG3}) for two different initial conditions at $N=3$ in the random case $n=0$.
    }
    \label{flows}
\end{figure}

The RG equations of the disordered case are given by (\ref{RG1})-(\ref{RG3}) with $n=0$. It is possible to check that they amount to those obtained for $N=2$ in \cite{DD_AT,Dotsenko_AT}, for $N$ generic in \cite{Murthy}, and for the $O(N)$-invariant case $\lambda_1=\lambda_3$ in \cite{Cardy_GN}. These works considered the RG flows ending in the pure Ising fixed point $\lambda_1=\lambda_2=\lambda_3=0$, and for $N>2$ this is sufficient to infer a softening of the first order transition of the pure model into a continuous transition with Ising exponents, and to ensure consistence with softening of first order transitions by disorder in two dimensions \cite{AW}. On the other hand, it was observed in \cite{random_line} that the pattern of softening in the $N$-color Ashkin-Teller model is richer. Indeed, the r.h.s. of (\ref{RG1})-(\ref{RG3}) with $n=0$ more generally vanish for 
\EQ
\lambda_1=0\,,\hspace{1cm}\lambda_2=\lambda_3\,,
\label{pert_line}
\EN
thus yielding a line of random fixed points for fixed $N$ which we know will persist to all orders, since (\ref{pert_line}) is the weak disorder limit of the exact scattering solution L1$_-$ with the identifications $Y_7\to\lambda_1$, $Y_1\to\lambda_2$ and $Y_4\to\lambda_3$. We show in Figure~\ref{flows} RG flows determined by (\ref{RG1})-(\ref{RG3}) with two different initial conditions at $n=0$, $N=3$. In the left panel the flow ends into the fixed point $\lambda_3=0$ on the line (\ref{pert_line}), namely the fixed point of the pure model. In the right panel, instead, the couplings flow towards a fixed point with $\lambda_3\neq 0$ on the line (\ref{pert_line}) and remain close to it for a substantial amount of the RG ``time" before moving away from it and eventually leaving the perturbative region $|\lambda_\alpha|\ll 1$ for values of $t$ not shown in the figure. The range of $t$ in which the second type of flow stays close to the random fixed point (\ref{pert_line}) with $\lambda_3\neq 0$ can be made arbitrarily large reducing the initial values of the couplings, which amounts to going deeper into the perturbative region and increasing the accuracy of the one-loop approximation. This is how the one-loop approximation indicates that the nonperturbative counterpart of the line of random fixed points (\ref{pert_line}) -- namely solution L1$_-$ -- is infrared stable for suitable choices of the system parameters. A line of fixed points at fixed $N$ allows for critical exponents continuously varying with the system parameters, thus explaining why these have been numerically observed in \cite{WD} for $N=2$ and in \cite{Katzgraber1,Katzgraber2} for $N=3$. This is not in conflict with the observation of Ising exponents at $N=3,4$ in \cite{Vojta_3color,Vojta_4color}, since we know that Ising criticality can also be realized for some choices of system parameters as the particular case $S_1=0$ of solution L1$_-$. In addition, as we will see in section~\ref{superuniversality}, the situation is even subtler, since some exponents indeed keep Ising values along the line of fixed points. 

Hence, we see that for $N>2$ weak disorder softens the first order transition of the pure model into a second order one, and that the large distance limit allows for the line of fixed points L1$_-$ at fixed $N$. Theoretically, the softening raises the question of the ultraviolet fixed point the RG flow ending in L1$_-$ originates from. Since the softening occurs for $N>2$, the ultraviolet fixed point should exist in the same range. In addition, it should be contained in Table~\ref{dis_sol} which lists the fixed points allowed by the symmetries of the model. It is then no coincidence\footnote{See \cite{colloquium,DL_softening} for an analogous phenomenon in the two-dimensional disordered $q$-state Potts model, where the softening occurs for $q>4$.} that Table~\ref{dis_sol} includes the solutions W1 and W2 defined for $N\geq 2$, with disorder vanishing at $N=2$. The main difference between the two solutions appears to be that the two disorder strength parameters $|S_4|$ and $|S_7|$ coincide in W2 while the second vanishes in W1. 

It is worth stressing why L1$_-$ is the only solution with weak disorder limit which has a perturbative counterpart in the fixed points of Eqs.~(\ref{RG1})-(\ref{RG3}). The reason is that the pure limit of L1$_-$ is P1$_-$, which is the fixed point around which Eqs.~(\ref{RG1})-(\ref{RG3}) have been derived. Instead, as we saw, the pure limit of W1, W2 and W3$^-$ is P5$_-$ with $Y_1=0$. Concerning W3$^+$ and W4$_-$, whose pure limit is at $N=1$, W4$_-$ belongs to the $O(N)$-invariant subspace and its pure limit is on the loop gas solution P2$_-$ \cite{colloquium,DL_ON1,DL_ON2} and has been studied perturbatively in \cite{Shimada}. The interpretation of the limit $N\to 1$ of W3$^+$ is less clear, also because $N=1$ is an endpoint of this solution. We saw that W3$^+$ and W3$^-$ are actually continuation of each other and must be regarded as a single solution; the only nontrivial fixed point they contribute for $N$ integer corresponds to W3$^+$ with $N=2$. Putting all together, we see that L1$_-$ is the only solution with weak disorder limit which is expected to play a role for $N$ generic in the description of the large distance critical properties of the model (\ref{lattice}).

\subsection{Solutions without weak disorder limit}
The solutions without weak disorder limit are those for which the two disorder strength parameters $|S_4|$ and $|S_7|$ do not simultaneously vanish at some value of $N$. As a matter of fact, all such solutions (L2, V-type and M-type in Table~\ref{dis_sol}) have $|S_7|=1$, namely the maximal value of $|S_7|$ allowed by (\ref{uni10}). Hence, we group these solutions according to the behavior of $|S_4|$. 

The solutions of type M are characterized by the fact that also the second disorder strength parameter takes its maximal value $|S_4|=1$. Analytical expressions for M1, M2 and M3 can be obtained but are cumbersome. As can be seen in Figure~\ref{fig:ampiezze_groups}, these solutions have merging points at\footnote{These two numbers are roots of the polynomial $x^7-12 x^6+51 x^5-17 x^4-157 x^3+78 x^2+113 x-65$.} $N=0.86525..$ and $N=2.1830..$ through which they combine nontrivially in what can be considered a single solution extending for all values of $N$. Solutions M4 and M5 coincide at $N=1$, which is the endpoint of the first and the starting point of the second. M6 is the only solution originating at $N>2$: it extends from $N=3.8284..$, where it coincides with M1, to infinity (Figure~\ref{fig:ampiezze_groups}). Solutions M7 and M8 are defined for any $N$ and are $N$-independent. M7 and M9 fall in the $O(N)$-invariant subspace (\ref{ON}) and were discussed in the study of the $O(N)$ model in \cite{colloquium,DL_ON2}.

\begin{figure}[t]
\centering
\includegraphics[width=6cm]{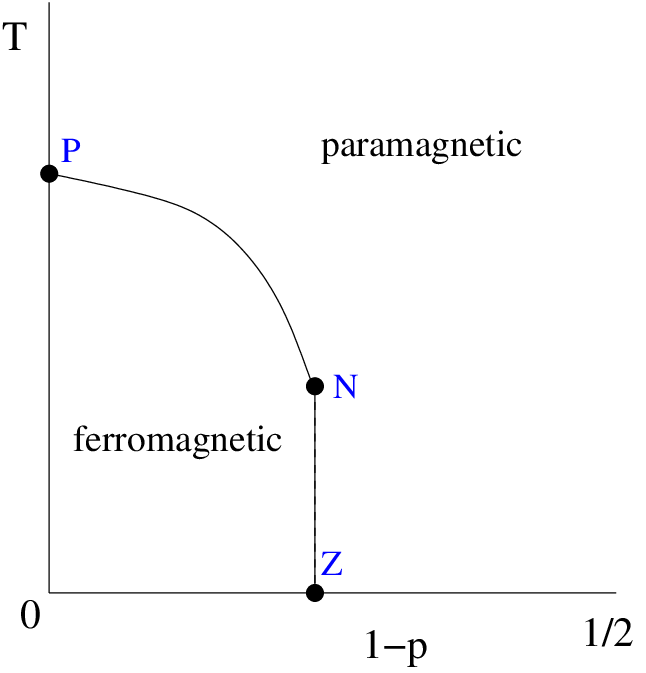}
\caption{The phase diagram of the the square lattice $\pm J$ random bond Ising model is symmetric under $p\to 1-p$. P, N and Z denote the pure, Nishimori and zero-temperature fixed points, respectively.}
\label{ising_pd}
\end{figure}

Insight into the meaning of the solutions of type M can be obtained starting from the Ising case ($N=1$), which has a single random coupling. Figure~\ref{ising_pd} shows the Ising phase diagram \cite{McMillan,MC,PHP,HPtPV} in two dimensions for the disorder distribution 
\EQ
P(J_{xy})=p\,\delta(J_{xy}-1)+(1-p)\,\delta(J_{xy}+1)\,,
\label{bimodal}
\EN
where the fraction $1-p$ of antiferromagnetic bonds is the disorder strength. There are three magnetic RG fixed points located along the ferromagnetic-paramagnetic phase boundary, namely the fixed point P of the pure model, the Nishimori multicritical point N \cite{GD_Nishimori,Nishimori,Nishimori_book,LdH1}, and the zero-temperature fixed point Z. It is known \cite{colloquium,DL_ON2} that the fixed points N and Z have $|S_0|=\sqrt{2}$ and $|S_7|=1$, and then maximize the value of the single disorder strength parameter $|S_7|$ ($S_4$ is not defined at $N=1$). This matches the exact lattice result that the Nishimori point N has maximal disorder on the phase boundary and that the fixed point Z can also have the same maximal disorder \cite{Nishimori_book}. Numerical simulations on the square lattice (see \cite{PHP} and references therein) show that Z occurs at a value of $1-p$ very slightly smaller than that of N, but this small re-entrance of the phase boundary is necessarily nonuniversal (i.e. lattice-dependent) and cannot show up in our field theoretical framework. This is why N and Z both maximize $|S_7|$. At the same time, we must consider that, while critical behavior at weak disorder does not depend on the distribution of a single random coupling (see e.g. \cite{Cardy_book}), this is not in general the case for stronger disorder. For example, for the randomly diluted Ising ferromagnet, corresponding to the disorder distribution
\EQ
P(J_{xy})=p\,\delta(J_{xy}-1)+(1-p)\,\delta(J_{xy})\,,
\label{dilution}
\EN
the phase diagram is qualitatively similar to that of Figure~\ref{ising_pd} but the multicritical point N is absent and Z becomes a percolation fixed point \cite{Cardy_book}. Also this percolation fixed point has $|S_0|=\sqrt{2}$ and $|S_7|=1$ \cite{colloquium}. 

With the exception of M6, which starts at $N=3.8284..$, all solutions of type M in Table~\ref{dis_sol} have $|S_0|=\sqrt{2}$ and $|S_7|=1$ at $N=1$ and can be considered as continuation in $N$ of the multicritical or zero-temperature Ising fixed points. Their relatively large number accounts for the fact that for $N>1$ there are two disorder strength parameters and that the combinations of two disorder distributions increase the possibilities of different critical behavior.

\begin{figure}[t]
    \centering
    \includegraphics[width=0.65\linewidth]{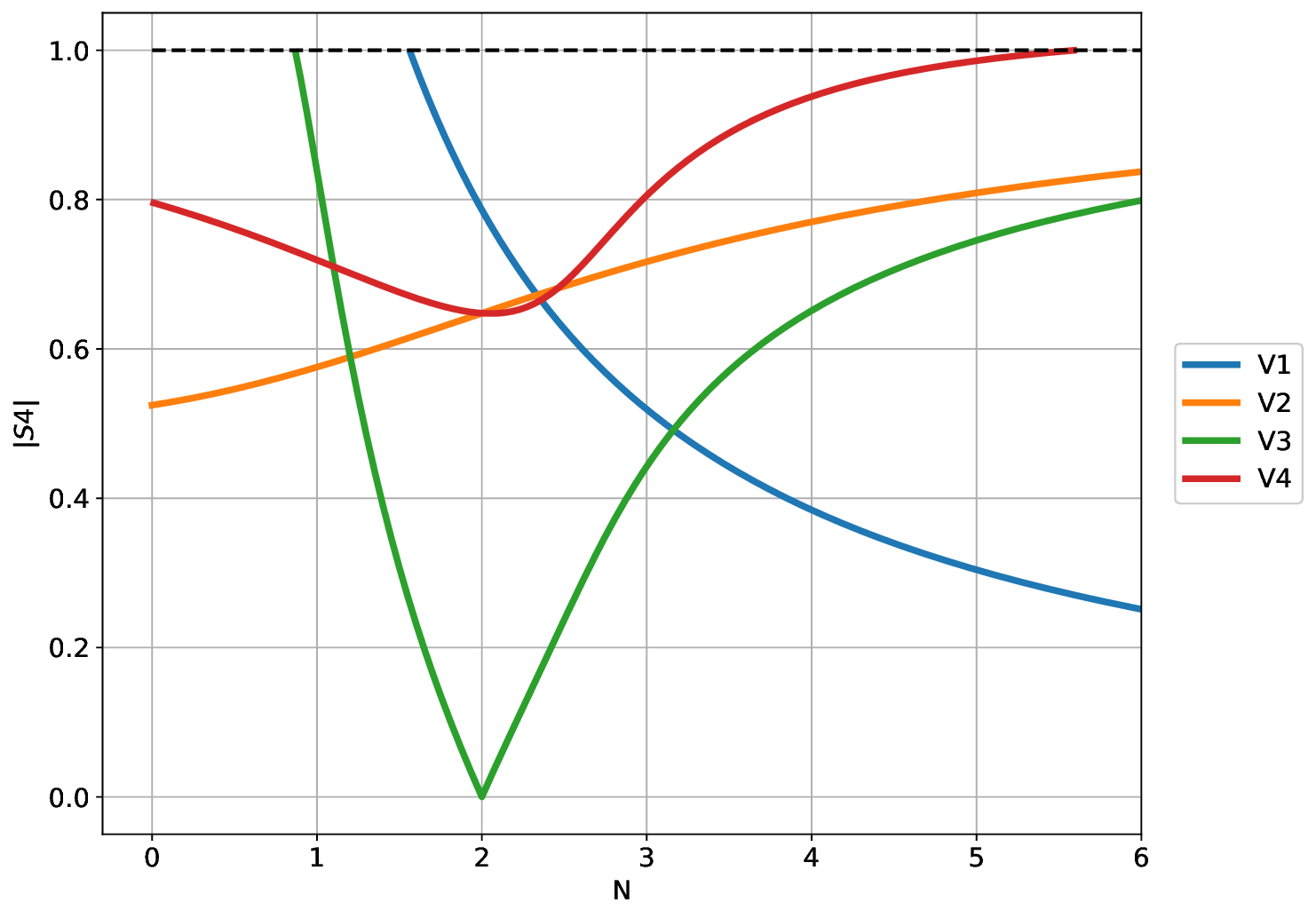}
    \caption{$|S_4|$ for the solutions V1-V4. The second disorder strength parameter takes its maximal value $|S_7|=1$.}
    \label{s4_v}
\end{figure}

The solutions of type V differ from those of type M for the fact that the second disorder strength parameter $|S_4|$ is not fixed to its maximal value. For V1-V4 it varies with $N$ (Figure~\ref{s4_v}), so that these solutions are allowed to merge solutions of type M when $|S_4|$ reaches its maximal value 1. And indeed, V1 at its starting value $N=(\sqrt{17}-1)/2$ coincides with M5 (which has $S_1=0$ at that point), V2 at its starting value $N=-2.4142..$ (possibly equal to $-\sqrt{2}-1$) coincides with M3, V3 at its starting value $N=0.86730..$ coincides with M1, while V4 terminates on M1 at $N=5.5986..$\,. Solution V5 has $S_1=S_4=0$ and corresponds to $N$ decoupled disordered Ising fixed points.

The solution L2 is defined only for $N=2$ but corresponds to a line of fixed points parametrized by $Y_4=\textrm{Im}(S_4)$. It coincides with the solutions M5, V1 and V3 for $|Y4|=1$, $0.78615..$ and 0, respectively.

\section{Critical indices and superuniversality}
\label{superuniversality}
The solutions of Table~\ref{dis_sol} allow the determination of the scattering phase (\ref{S}) for the case $n=0$ of our interest, and then the determination -- through (\ref{phase}) -- of the conformal dimension $\Delta_\eta$ of the chiral field $\eta$ which creates the right-moving particles. Right-left symmetry ensures that the chiral field $\bar{\eta}$ which creates the left-moving particles has $\bar{\Delta}_{\bar{\eta}}=\Delta_\eta$. The scalar field $\eta\bar{\eta}$ has scaling dimension $X_{\eta\bar{\eta}}=2\Delta_\eta$, creates the state $\sum_{a,i}|a_ia_i\rangle$, and is one of the symmetry-invariant scalar fields $\varepsilon_k$ at the given fixed point. We order these fields according to their scaling dimension: $X_{\varepsilon_1}<X_{\varepsilon_2}<\cdots$. They determine, in particular, the correlation length critical exponents $\nu_k=1/(2-X_{\varepsilon_k})$. 

\begin{figure}[htbp]
\centering
\begin{subfigure}{\textwidth}
    \centering
    \includegraphics[width=\linewidth]{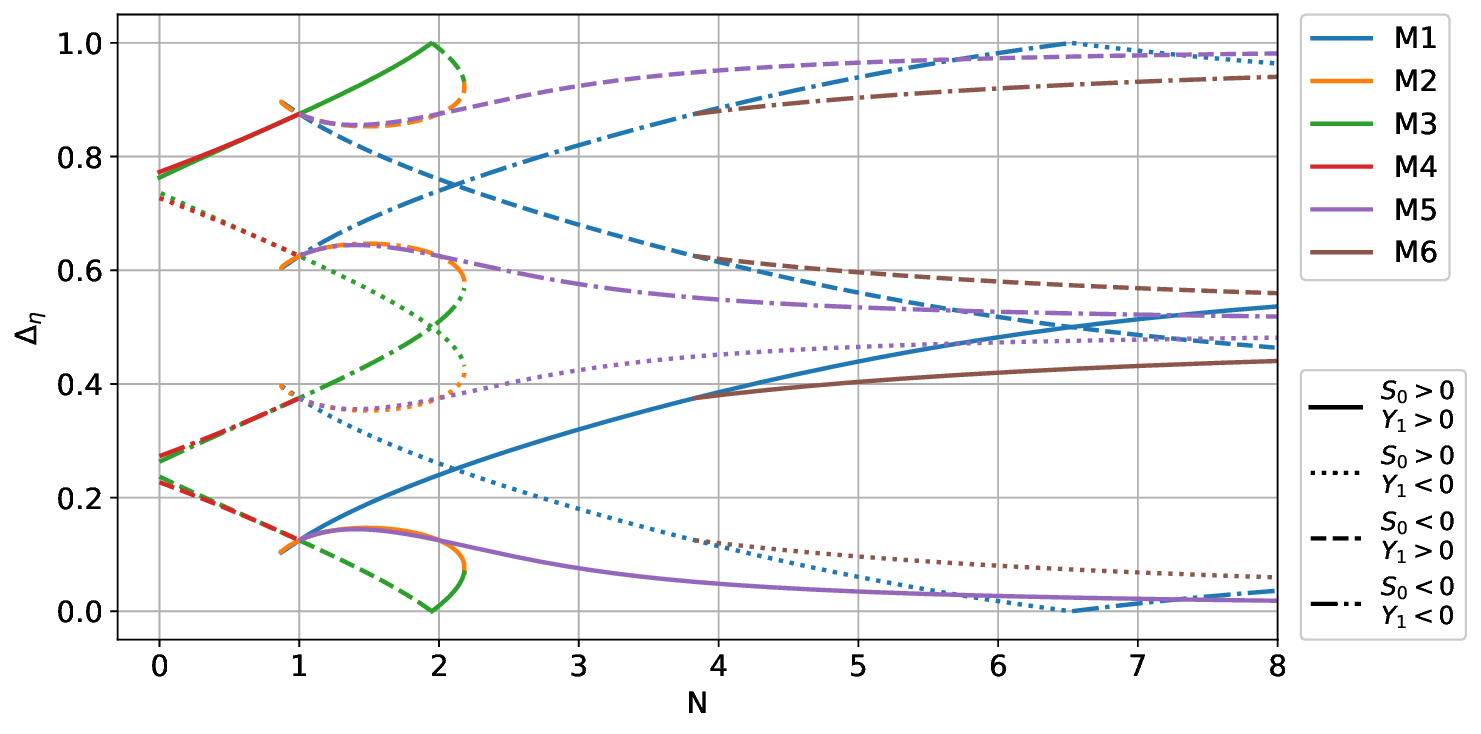}
     \caption{$ $}
\end{subfigure}
 \vspace{0.5cm} 
\begin{subfigure}{\textwidth}
    \centering
    \includegraphics[width=1\linewidth]{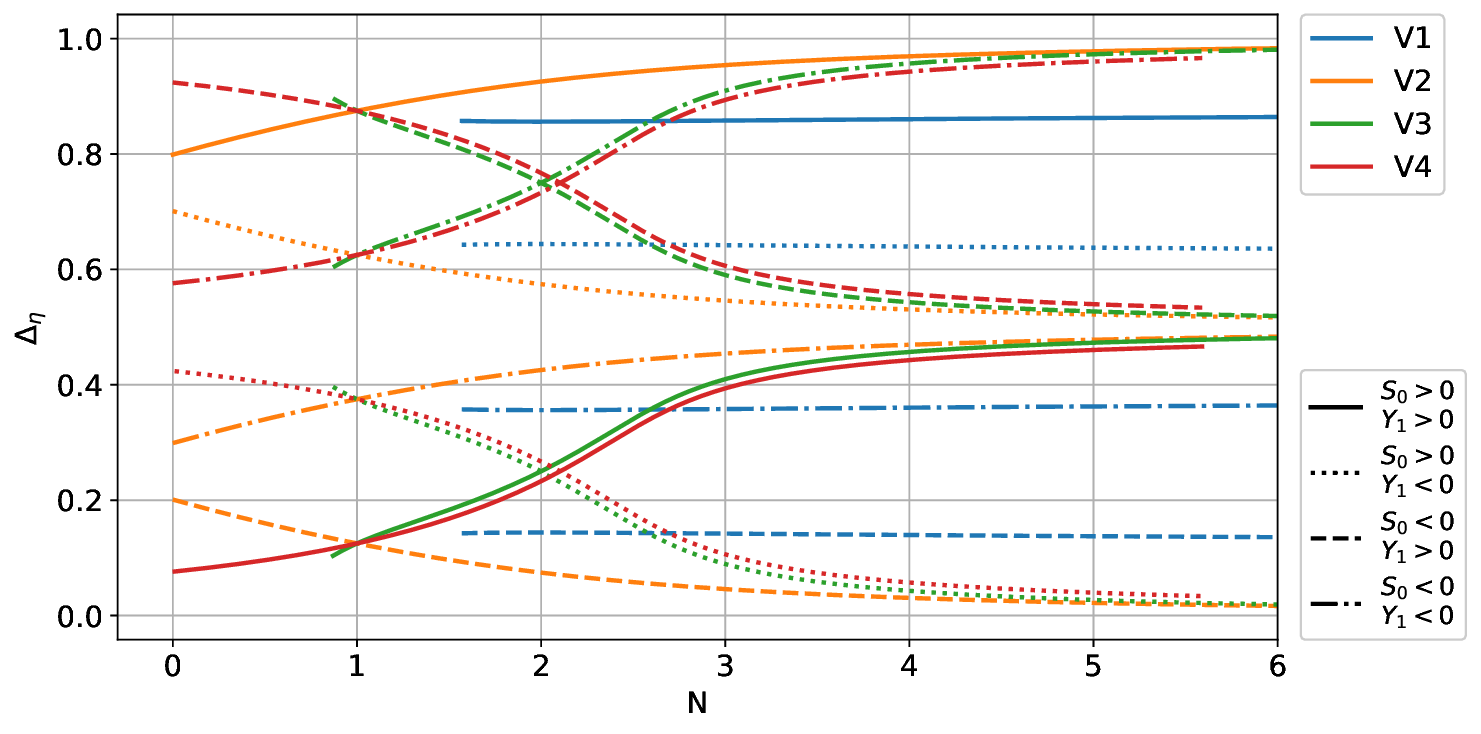}
    \caption{$ $}
\end{subfigure}
\caption{$\Delta_\eta$\,(mod\,1) for the solutions not given analytically in Table~\ref{dis_sol} (with the exception of V1). The amplitude signs in the legends refer to the lowest positive value of $N$ for which the solution is defined. Also for V1 the results are $N$-dependent.
}
\label{fig:chiraldim}
\end{figure}

The results for $\Delta_\eta$ for the solutions of Table~\ref{dis_sol} are shown in Table~\ref{table_S} and Figure~\ref{fig:chiraldim}. A particularly interesting feature of these results is that sometimes $\Delta_\eta$ is $N$-independent, even in cases when the corresponding solution depends on $N$. Since the global internal symmetry of the model varies with $N$ and, on universality grounds, critical exponents in fixed dimension are normally expected to depend precisely on the symmetry, this means that some solutions allow for ${\it some}$ critical exponents which are ``superuniversal". This exact analytical mechanism, which does not occur in pure systems, was first exhibited in \cite{random} for the two-dimensional random bond $q$-state Potts model, finally explaining the apparent -- i.e. within error bars -- $q$-independence of the correlation length exponent $\nu$ observed in \cite{CFL} and subsequent numerical studies at the fixed point with weakest disorder. It was observed in \cite{colloquium} that superuniversality of some critical exponents is not rare in random criticality and it was argued that it is not necessarily limited to two dimensions. This prediction was recently confirmed in \cite{GD_Nishimori} through the determination of exact critical exponents at the Nishimori multicritical point in two and three dimensions. 

\begin{table}[t]
    \centering
    \begin{tabular}{c|c|c}
    \hline 
         Solution & $S$ & $\Delta_\eta$  \\
         \hline\hline
         L2 & $ [\pm]\left( \frac{Y_4}{\sqrt{1 + Y_4^2}} \mp i\,\textrm{sgn}(S_7) \frac{1}{\sqrt{1+Y_4^2}} \right) $ & $\left[ \frac{1}{8}, \frac{3}{8} \right]$,$ \left[ \frac{5}{8}, \frac{7}{8} \right] $ \\
         L1, W1, W4 & $ \textrm{sgn}(S_0)$ & $0,\frac{1}{2}$ \\
         W2 & $ \textrm{sgn}(X_1)$ & $0,\frac{1}{2}$ \\ 
         W3$^\pm$ & $\pm \sqrt{1-(N^2-N)^2f^2_\pm(N)} [\pm] i (N^2-N)f_\pm(N)$ & $(i/2\pi)\ln S$\\
         V1 & $ \pm \sqrt{1 - \frac{N^2}{4} \frac{g^2(N) -2}{N-1} } [\pm] i \frac{N}{2} \sqrt{\frac{g^2(N) -2}{N-1}} $ & $(i/2\pi)\ln S$\\
         V5, M7 & $ \pm \frac{1}{\sqrt{2}} [\mp] i\frac{1}{\sqrt{2}} $ & $\frac{1}{8},\frac{3}{8},\frac{5}{8},\frac{7}{8}$\\
         M8 & $ \pm \frac{1}{\sqrt{2}} [\pm] i\frac{1}{\sqrt{2}} $ & $\frac{1}{8},\frac{3}{8},\frac{5}{8},\frac{7}{8}$\\
         M9 & $ \frac{\sqrt{ 2 }}{2(N^{2}+1)}(\pm (-N^{2}+2N+1) [\pm] i(N^{2}+2N-1))$ & $(i/2\pi)\ln S$\\
         \hline 
    \end{tabular}
    \caption{Scattering phase (\ref{S}) and values of $\Delta_\eta$\,(mod\,1) for the analytical solutions of Table~\ref{dis_sol}. Sign conventions correspond to those of Table~\ref{dis_sol}. Values (intervals for L2) of $\Delta_\eta$ separated by commas correspond to different sign choices.}
    \label{table_S}
\end{table}

In the present case of the $N$-color Ashkin-Teller model, we know that L1$_-$ and the solutions of type W possess limits of vanishing disorder where they reduce to free fermions and have $\Delta_\eta=1/2$. We now see that for L1$_-$, W1$_-$, W2 and W4$_-$ this value $\Delta_\eta=1/2$ is superuniversal, i.e. $N$-independent. For L1$_-$, in particular, this means that the scaling dimension of the energy density field $\varepsilon=\varepsilon_1=\eta\bar{\eta}$ keeps its Ising value 1 for all $N$, but also for all values of the parameter $Y_1$ which spans the line of fixed points for a given $N$ larger than 1. On the other hand, the scaling dimension $X_\sigma$ of the spin fields can vary continuously with $Y_1$. The consequent pattern of fixed Ising thermal critical exponents $\nu=1$ and $\alpha=0$, and continuously varying magnetic exponents $\beta$ and $\gamma$, explains the numerical results obtained for $N=2$ in \cite{WD}. It also sheds light on the observation for $N=3,4$ of continuously varying exponents in \cite{Katzgraber1,Katzgraber2} and of Ising exponents in \cite{Vojta_3color,Vojta_4color}.

The other solutions with superuniversal $\Delta_\eta$ are\footnote{The case of V5 is trivial, since it corresponds to $N$ decoupled Ising fixed points.} M7 and M8, which actually are completely $N$-independent and yield values of $\Delta_\eta$ which are odd integer multiples of $1/8$. We saw that these solutions can be seen as continuations in $N$ of the Ising zero-temperature and Nishimori fixed points at $N=1$. The Ising Nishimori multicritical point possesses two relevant fields $\varepsilon_1$ and $\varepsilon_2$ with scaling dimensions $X_{\varepsilon_1}=4/3$ and $X_{\varepsilon_2}=7/4$ \cite{GD_Nishimori}, from which we deduce $\eta\bar{\eta}=\varepsilon_2$ ($\Delta_\eta=7/8$). The superuniversality of $X_{\varepsilon_2}$ within the solutions M7 and M8 is similar to that found in \cite{GD_Nishimori} for the Nishimori point in the three-dimensional $O(N)$ model.

We recalled in the previous section that in the randomly diluted Ising ferromagnet the zero temperature fixed point is a percolation point. Two-dimensional percolation has $X_{\varepsilon_1}=5/4$ \cite{Nienhuis}, from which we deduce $\eta\bar{\eta}=\varepsilon_1$ ($\Delta_\eta=5/8$). The fact that this percolation fixed point can extend to any $N$ can be understood observing that in the randomly dilute ferromagnet the zero-temperature transition to the paramagnetic phase corresponds to percolation of a cluster of empty sites, and that the color of the clusters of occupied sites is not important.

\section{Conclusion}
\label{conclusion}
In this paper we considered the two-dimensional $N$-color Ashkin-Teller model, which amounts to $N$ Ising models coupled by four-spin interactions of energy-energy type. More specifically, we have been interested in the critical properties of the model in presence of quenched disorder. The model is known to be interesting already in absence of disorder, since it yields continuously varying exponents for $N=2$ and, for $N>2$, a first order transition produced by a RG mechanism which generates a finite correlation length at the transition point. 

A characteristic feature of the disordered model is that it allows for two independent disorder strength parameters, one associated to the interaction within each Ising copy and the other associated to the interaction between different copies (i.e. different colors). This possibility of combining two disorder distributions can then be expected to produce a richer pattern of random critical behavior than that allowed in models such as $O(N)$ or $q$-state Potts which possess a single disorder strength parameter. This expectation has been confirmed by our study within the scattering framework, through which we determined all the solutions of the exact RG fixed point equations. In addition, we showed that the number of fixed points allowed by the symmetries of the model increases when moving from weak to strong disorder, a finding which again has a clear interpretation. Indeed, it can be generally shown on perturbative grounds that weak relevant disorder leads to a new fixed point which does not depend on the distribution of a single random coupling \cite{Cardy_book}. This is why, for example, weak random frustration and weak random dilution yield the same critical exponents in the three-dimensional Ising model. On the other hand, these two disorder distribution produce different critical behavior when disorder becomes stronger. 

We showed that the weak disorder sector of the two-dimensional $N$-color Ashkin-Teller model is characterized by the presence of an exact solution that, for $N$ fixed and larger than 1, corresponds to a line of fixed points along which the magnetic critical exponents vary continuously while the thermal exponents keep Ising values. This peculiar pattern explains the numerical results of \cite{WD} at $N=2$. For $N=3$ it was numerically captured in part in \cite{Katzgraber1,Katzgraber2} through the observation of continuously varying exponents, and in part in \cite{Vojta_3color} through the observation of Ising exponents. On the other hand, the fixed point solutions with maximal disorder extending down to $N=1$ can be seen as continuations in $N$ of the Nishimori and zero-temperature fixed points of the Ising model. Our result that outside $N=1$ they split into several branches reflects the passage from one to two random couplings. 

Also for this model we found that some fixed point solutions possess some critical exponents which do not depend on the symmetry parameter ($N$ in the present case) and are in this sense ``superuniversal", further confirming the observation that this remarkable phenomenon is a rather common feature of random criticality \cite{colloquium}. We also showed how the exact scaling dimension directly readable from the scattering matrix matches the value known from other exact sources in the two cases of random criticality (Nishimori multicritical point \cite{GD_Nishimori} and percolation \cite{Nienhuis}) for which these other sources are available.

\end{document}